\documentclass[prd,twocolumn,aps,showpacs]{revtex4}

\def\[{\left\lbrack}
\def\]{\right\rbrack}

\def\({\left(}
\def\){\right)}
\newcommand{\be}{\begin{equation}}
\newcommand{\ee}{\end{equation}}
\newcommand{\ea}{\end{eqnarray}}
\newcommand{\ba}{\begin{eqnarray}}

\newcommand{\ep}{{\epsilon}}

\def\ep{\epsilon}

\begin{document}

 \title{Duality through the symplectic embedding formalism}

\author{E. M. C. Abreu$^a$\footnote{\sf E-mail: evertonabreu@ufrrj.br},A.C.R. Mendes$^b$\footnote{\sf E-mail: albert@fisica.ufjf.br},C. Neves$^b$\footnote{\sf E-mail: cneves@fisica.ufjf.br},W. Oliveira$^b$\footnote{\sf E-mail: wilson@fisica.ufjf.br} and F.I. Takakura$^b$\footnote{\sf E-mail: takakura@fisica.ufjf.br}} 
\affiliation{${}^{a}$Departamento de F\'{\i}sica, Universidade Federal Rural do Rio de Janeiro\\
BR 465-07, 23851-180, Serop\'edica, Rio de Janeiro, Brazil\\
${}^{b}$Departamento de F\'{\i}sica,
ICE, Universidade Federal de Juiz de Fora,\\
36036-330, Juiz de Fora, MG, Brazil\\
\bigskip
\today}


\begin{abstract}
\noindent  In this work we show that we can obtain dual equivalent actions following the symplectic formalism with the introduction of extra variables which enlarge the phase space.  We show that the results are equal as the one obtained with the recently developed gauging iterative Noether dualization method (NDM). We believe that, with the arbitrariness property of the zero mode, the symplectic embedding method (SEM) is more profound since it can reveal a whole family of dual equivalent actions.
We illustrate the method demonstrating that the gauge-invariance of the electromagnetic Maxwell
Lagrangian broken by the introduction of an explicit mass term and a
topological term  can be restored to obtain the dual equivalent and
gauge-invariant version of the theory.  
\end{abstract}

\maketitle

\section{Introduction}

Duality is an useful concept in field theory and statistical mechanics since there are very few analytic tools available for studying non-perturbative properties of systems with many degrees of freedom.

Recently, the so-called gauging iterative Noether dualization method \cite{ainrw} has 
been shown to thrive in establishing some dualities between models \cite{iw,iw2,iw3}.
This method is based on the traditional concept of a local lifting of a global symmetry and may be
realized by an iterative embedding of Noether counterterms.  However, this
method provides a strong suggestion of duality since it has been shown to give
the expected result in the paradigmatic duality between the so-called self-dual model \cite{tpn} and the Maxwell-Chern-Simons 
theory in three dimensions duality.  This correspondence was first established by Deser and Jackiw \cite{dj} 
and using a parent action approach \cite{suecos}.

At the same time, we know that in the literature there are several schemes to reformulate noninvariant models as gauge theories.  Some constraint-conversion formalisms, based on Dirac's method \cite{dirac,dirac2,dirac3,dirac4,dirac5}, were developed following Faddeev's idea of phase-space extension with the introduction of auxiliary variables \cite{FS}.  Among them, the BFFT \cite{batalin,batalin2} and the iterative \cite{wotzasek,wotzasek2} methods were powerful enough to be successfully applied to a great number of important physical models.  Although these techniques share the same conceptual basis \cite{FS} and follow Dirac's framework \cite{dirac,dirac2,dirac3,dirac4,dirac5}, these constraint-conversion methods were implemented following different directions.  Historically, both BFFT and the iterative methods were to deal with linear systems such as chiral gauge theories \cite{wotzasek,wotzasek2,mo,mo2} in order to eliminate the gauge anomaly that hampers the quantization process.  In spite of the great success achieved by these methods, they have an ambiguity problem \cite{barcelos}.  This problem naturally arises when the second-class constraint is converted into a first-class one with the introduction of WZ variables.  Due to this, the constraint conversion process may become a hard task. \cite{barcelos}.

The symplectic embedding method  \cite{aon} is not affected by this ambiguity problem.  It has the great advantage of being a simple and direct way of choosing the infinitesimal gauge generators of the built gauge theory.  This give us a freedom to choose the content of the embedded symmetry according to our necessities.  This feature makes possible a greater control over the final Lagrangian.  This method can avoid the introduction of infinite terms in the Hamiltonian of embedded non-commutative and non-Abelian theories.  This can be accomplished because the infinitesimal gauge generators are not deduced from previous unclear choices.  Another related advantage is the possibility of doing a kind of general embedding, that is, instead of choosing the gauge generators at the beginning, one can leave some unfixed parameters with the aim of fixing them later, when the final Lagrangian has being achieved.  Although one can reach faster the final theory fixing such parameters as soon as possible, this path is more interesting in order to study the considered theory, and it is helpful if the desired symmetry is unknown, but some aspects of the Lagrangian are wanted.  

We should mention that this approach to embedding is not dependent on any undetermined constraint structure and also works for unconstrained systems.  This is different from all the existent embedding techniques that we use to convert \cite{batalin,batalin2,wotzasek,wotzasek2}, project \cite{vyth} or reorder \cite{mitra} the existent second-class constraints into a first-class system.  This technique on the other hand only deals with the symplectic structure of the theory so that the embedding structure does not rely on any pre-existent constrained structure.

In \cite{aon} two of us demonstrated that the SEM does not change the physical contents originally present in the theory computing the energy spectrum.  This technique follows Faddeev's suggestion \cite{FS} and is set up on a contemporary framework to handle noninvariant models, namely, the symplectic formalism \cite{FJ,FJ2,FJ3,FJ4}.

The purpose of the present paper is to study a version of the Maxwell Lagrangian density modified by the introduction of an explicit mass term and of a topological term \cite{Carroll,helayel}.

In \cite{Carroll} Carroll, Field and Jackiw analyzed the consequences of introducing an explicit mass term and a topological term in the electromagnetic Maxwell Lagrangian.  The action introduced was, 
\be\label{0.1}{\cal L} \[A_\mu \] = -{\beta \over
4}F_{\mu\nu}F^{\mu\nu} +{m^2\over 2}A_\mu A^\mu \,+\,L_{CS}\[A_\mu \],
\ee
where $L_{CS}$ is a four-dimensional version of the Chern-Simons (CS) action, which couples the dual electromagnetic tensor to an external four-vector $p$,
\be \label{0.2} 
{\cal L}_{CS} \[A_\mu \] =-{1\over 4}p_\alpha
A_\beta \epsilon^{\alpha\beta\mu\nu}F_{\mu\nu}. 
\ee
This modification, which couples the electromagnetic field to a
``constant" external four-vector $p_\alpha$,  violates the Lorentz
invariance and parity, while preserving gauge invariance.

Some authors have explored the physical aspects of this model \cite{soldati,klink,ggw}.
As a theory for a modified electromagnetism it has been shown that the vacuum becomes a
birefringent media, and it was realized that this effect could be used to set limits
in the magnitude of the Lorentz violating vector $p_{\mu}$. In \cite{Carroll} only a
time-like $p_{\mu}$ was considered and it was argued that astrophysical observations
of polarized light and geomagnetic data seems to rule out a non-vanishing magnitude
of $p_{\mu}$ in this case. For the space-like case, astronomical observations
\cite{nodland} were used to argue in favor of a non-vanishing value of the magnitude
of $p_{\mu}$ but the results has been disputed \cite{cohen}.

Discussions concerning the consistence of the quantum field theory (QFT) defined by
(\ref{0.1}) as a function of the Lorentz character of $p_{\mu}$ has also been carried
out \cite{soldati, klink}. It was noted that a time-like $p_{\mu}$ gives rise to a
QFT for which unitarity and microcausality cannot be satisfied simultaneously. On the
other hand it seems that a consistent QFT can be defined for a space-like $p_{\mu}$.

So, the construction of dual equivalent and a gauge-invariant version of the Maxwell modified
theory, Eq.(\ref{0.1}), will be accomplished in the symplectic framework.
The SEM introduces extra variables which enlarge the
phase space \cite{aon} furnishing a dual equivalent action of the first one, and, furthermore restore the
gauge-invariance of the theory.

It is important to notice that more than one WZ symmetry will be unveiled (see \cite{fluidos} for a review), showing that the studied model does not have a unique WZ gauge-invariant description \cite{henrique}, but a family of dynamically equivalent WZ gauge-invariant representations.  This can allow an interesting discussion concerning both obvious symmetry (phase symmetry) and hidden symmetry (Galileo antiboost invariance) of the studied model.   For example, the additional symmetries found in \cite{bazeia} were investigated in \cite{fluidos} from the symplectic embedding point of view.  Indeed, the global status of these symmetries will be lifted to local.  We believe that this property of unveiling a whole family of symmetries and consequently a whole family of equivalent actions is the great advantage of the SEM in comparison with NDM.

In order for this work to be self-sustained, it is organized as
follows: In section II, we review the main steps, using the NDM, to obtain the dual equivalent action to the Maxwell-Chern-Simons theory \cite{helayel}.  In section III, we present a brief review of the
symplectic embedding formalism. In section IV, the
Maxwell-Chern-Simons (MCS) theory will be analyzed  from the
symplectic point of view \cite{FJ,FJ2,FJ3,FJ4}. Here, the Dirac brackets
among the fields will be computed. In section V, the SEM will be used and, as a consequence, the
gauge-invariant/dual equivalent version of the MCS theory will be obtained. In
the last section, we present our concluding observations and final
comments.

\section{The Noether dualization method}

In recent papers \cite{iw,iw2,iw3,ainrw,helayel} an alternative way to establish dual equivalences between gauge and non-gauge theories has been proposed that is based on the local lifting of the global symmetries present in the gauge action.  This is accomplished by iteratively incorporating counter-terms into the action depending on powers of the so-called Euler vectors.  These last ones are defined by the independent variations of the action, whose kernel gives the equations of motion, along with a set of auxiliary fields.  The resulting embedded theory is dynamically equivalent to the original one.  This gauge embedding approach has been applied to the case of the three-dimensional non-Abelian self-dual action which was proved to be equivalent to the Yang-Mills-Chern-Simons theory for the full range of values of the coupling constant and not only on the weak coupling regime.
This alternative approach to dual transformation that is dimensionally independent and sufficiently general to embody both Abelian and non-Abelian symmetries is used to analyze the dual equivalence of certain actions with dynamical couplings with both fermionic and bosonic matter fields.

Now we will follow the main steps in \cite{helayel} concerning the dualization of the massive Maxwell-Chern-Simons Lagrangian in four dimensions through NDM.  The MCS action is given by,
\be 
\label{e1} L^{(0)}={- \beta \over 4} F_{\mu
\nu}F^{\mu \nu}+
\,\frac{m^2}{2}A_{\mu}A^{\mu}\,-\,\frac{1}{4}p_{\alpha}A_{\beta}\ep^{\alpha
\beta \mu\nu}F_{\mu \nu}. 
\ee 
and the respective gauge symmetry is, 
\be 
A_{\alpha}\to
A_{\alpha}\,+\,\partial_{\alpha}\eta , ~~~~~~~~~\delta A_{\alpha}=\partial_{\alpha}\eta \,\,. 
\ee 
With the first variation of this Lagrangian, 
\be 
\delta L^{(0)}[A_{\mu}]\,=\,\left( m^2 A_{\mu}+ \beta (\partial ^\nu F_{\nu \mu}) 
-\ep_{\alpha \beta \nu \mu}p^\alpha (\partial^\beta A^\nu)\right)\partial^\mu \eta\,\,, 
\ee 
the Noether current can be computed as 
\be 
\label{e2} J_{\mu}\,=\,m^2 A_{\mu}+ \beta (\partial
^\nu F_{\nu \mu})- \ep_{\alpha \beta \nu \mu}p^\alpha
(\partial^\beta A^\nu)\,\,, 
\ee 
and the first iterated Lagrangian was obtained introducing an auxiliary field $B$,
$L^{(1)}\,=\,L^{(0)}\,-\,JB$. Following the NDM procedure we see that the $B$ field 
transforms as,
\be 
\delta \, B_{\mu}\,=\,\delta\,A_{\mu} =
\partial_{\mu} \eta,
\ee then \be \label{e3}
\delta\,L^{(1)}\,=\,-(\delta\,J_{\mu})\,B^{\mu}. 
\ee 
We have also that,
\be \label{e4}
\delta\,J_{\mu}\,=\,m^2\delta\,A_{\mu}\,=\,m^2(\partial_{\mu}
\eta). 
\ee 
Substituting this back, we have that the second iterated Lagrangian is,
\be
 L^{(2)}\,=\,L^{(1)}\,+\,\frac{m^2}{2}\,B_{\mu}B^\mu \,\,.
\ee
Using (\ref{e3}) and (\ref{e4}), the total
variation vanishes, i.e., $\delta L^{(2)}\,=\,0$, and 
the final form of this action is,
\ba \label{e5}
L^{(2)}\,&=&{- \beta \over 4} F_{\mu \nu}F^{\mu
\nu}+\,\frac{m^2}{2}A_{\mu}A^{\mu}\, -
\,\frac{1}{4}p_{\alpha}A_{\beta}\ep^{\alpha\beta
\mu\nu}F_{\mu
\nu}  \nonumber  \\
&-&\left( m^2 A_{\mu}+ \beta (\partial ^\nu F_{\nu \mu})-
\ep_{\alpha \beta \nu \mu}p^\alpha (\partial^\beta A^\nu)
 \right)\,B^{\mu}\, \nonumber  \\
&+&\,\frac{m^2}{2}\,B_\mu B^\mu \,\,,
\ea
where, solving for $B$ the equation of motion is,
\be
-J + m^2 B=0\,\,.
\ee
Substituting this result in (\ref{e5}), we obtain the gauge invariant/dual equivalent theory,
\ba
\label{e6}
L&=& {\beta \over 4 } F_{\mu \nu}F^{\mu \nu} +\frac 12 \ep_{\alpha \beta \nu \mu}p^\alpha (\partial^\beta A^\nu)A^ \mu \nonumber  \\
&-&\frac {1}{2m^2} \left ( \ep_{\alpha \beta \nu \mu}p^\alpha (\partial^\beta A^\nu)\right)^2 \\
&-& \frac {\beta ^2}{2m^2}(\partial _\mu F^{\mu \nu})^2 +\frac{\beta}{m^2}
\ep_{\alpha \beta \nu \mu}p^\alpha (\partial^\beta A^\nu)(\partial_ \rho F^{\rho \mu})\,\,. \nonumber  
\ea
We can see that in the limit $\beta=0$, the action (\ref{e6}) is a kind of 
parent action for this duality, this is a frequent behavior of NDM.
We can see it readily, noting that the last iterated Lagrangian (\ref{e5}) is
precisely the so-called parent action. Namely, by varying this with respect, first to $A$ and
second to $B$, the actions (\ref{e6}) and (\ref{e1}) are obtained \cite{helayel,br}.

This shows us that this duality is analogous to the one in three dimensions between
the Self-Dual model (SD) and the Maxwell-Chern-Simons (MCS). If we choose
the external vector $p_\mu$ to coincide with an (spacial) element of the space-time, 
and writing the fields in components, one may verify that the
action reduces to the SD action in {\it three} dimensions and the
action (\ref{e1}) coincides with the MCS one.  The duality is preserved in $D=4$.

\section{The symplectic embedding method}

In this section, we briefly review the symplectic embedding technique that
restore the gauge symmetry. This technique follows the Faddeev-Shatashivilli's
suggestion \cite{FS} and is set up on a contemporary framework to handle constrained
models, the symplectic formalism \cite{FJ,FJ2,FJ3,FJ4}.

In order to systematize the symplectic embedding formalism, we consider a general
noninvariant mechanical model whose dynamics is governed by a Lagrangian
${\cal L}(a_i,\dot a_i,t)$, (with $i=1,2,\dots,N$), where $a_i$ and $\dot a_i$
are the space and velocities
variables, respectively. Notice that this model does not result in the loss of
generality nor physical content. Following the symplectic method the zeroth-iterative
first-order Lagrangian one-form is written as
 \begin{equation}
\label{2000}
{\cal L}^{(0)}dt = A^{(0)}_\theta d\xi^{(0)\theta} - V^{(0)}(\xi)dt,
\end{equation}
and the symplectic variables are
\be
\xi^{(0)\theta} =  \left\{ \begin{array}{ll}
                               a_i, & \mbox{with $\theta=1,2,\dots,N $} \\
                               p_i, & \mbox{with $\theta=N + 1,N + 2,\dots,2N ,$}
                           \end{array}
                     \right.
\ee
where $A^{(0)}_\theta$ are the canonical momenta and $V^{(0)}$ is the symplectic potential. From the Euler-Lagrange equations of motion, the symplectic tensor is obtained as
\begin{eqnarray}
\label{2010}
f^{(0)}_{\theta\beta} = {\partial A^{(0)}_\beta\over \partial \xi^{(0)\theta}}
-{\partial A^{(0)}_\theta\over \partial \xi^{(0)\beta}}.
\end{eqnarray}
When the two-form $f \equiv \frac{1}{2}f_{\theta\beta}d\xi^\theta \wedge d\xi^\beta$ is singular, the symplectic matrix (\ref{2010}) has a zero-mode $(\nu^{(0)})$ that generates a new constraint when contracted with the gradient of the symplectic potential,
\begin{equation}
\label{2020}
\Omega^{(0)} = \nu^{(0)\theta}\frac{\partial V^{(0)}}{\partial\xi^{(0)\theta}}.
\end{equation}
This constraint is introduced into the zeroth-iterative Lagrangian one-form Eq.(\ref{2000}) through a Lagrange multiplier $\eta$, generating the next one
\begin{eqnarray}
\label{2030}
{\cal L}^{(1)}dt &=& A^{(0)}_\theta d\xi^{(0)\theta} + d\eta\Omega^{(0)}- V^{(0)}(\xi)dt,\nonumber\\
&=& A^{(1)}_\gamma d\xi^{(1)\gamma} - V^{(1)}(\xi)dt,\end{eqnarray}
with $\gamma=1,2,\dots,(2N + 1)$ and
\begin{eqnarray}
\label{2040}
V^{(1)}&=&V^{(0)}|_{\Omega^{(0)}= 0},\nonumber\\
\xi^{(1)_\gamma} &=& (\xi^{(0)\theta},\eta),\\
A^{(1)}_\gamma &=&(A^{(0)}_\theta, \Omega^{(0)}).\nonumber
\end{eqnarray}
As a consequence, the first-iterative symplectic tensor is computed as
\begin{eqnarray}
\label{2050}
f^{(1)}_{\gamma\beta} = {\partial A^{(1)}_\beta\over \partial \xi^{(1)\gamma}}
-{\partial A^{(1)}_\gamma\over \partial \xi^{(1)\beta}}.
\end{eqnarray}
If this tensor is nonsingular, the iterative process stops and the Dirac's brackets
 among the phase space variables are obtained from the inverse matrix
 $(f^{(1)}_{\gamma\beta})^{-1}$ and, consequently, the Hamilton equation of
 motion can be computed and solved, as discussed in \cite{gotay}. It is well known
 that a physical system can be described at least classically in terms of a symplectic
 manifold $M$. From a physical point of view, $M$ is the phase space of the system while
 a nondegenerate closed 2-form $f$ can be identified as being the Poisson bracket. The
 dynamics of the system is  determined just specifying a real-valued function (Hamiltonian)
$H$ on the phase space, {\it i.e.}, one of these real-valued function
solves the Hamilton equation, namely,
\be \label{2050a1}
\iota(X)f=dH, \ee
and the classical dynamical trajectories of the
system in the phase space are obtained. It is important to mention
that if $f$ is nondegenerate, Eq. (\ref{2050a1}) has an unique
solution. The nondegeneracy of $f$ means that the linear map
${\flat}:TM\rightarrow T^*M$ defined by ${\flat}(X):={\iota}(X)f$ is an
isomorphism, due to this, the Eq.(\ref{2050a1}) is solved uniquely
for any Hamiltonian $(X={\flat}^{-1}(dH))$. On the contrary, the
tensor has a zero-mode and a new constraint arises, indicating
that the iterative process goes on until the symplectic matrix
becomes nonsingular or singular. If this matrix is nonsingular,
the Dirac's brackets will be determined. In Ref. \cite{gotay}, the
authors consider in detail the case when $f$ is degenerate, which
usually arises when constraints are presented in the system. In
which case, $(M,f)$ is called the presymplectic manifold. As a
consequence, the Hamilton equation, Eq. (\ref{2050a1}), may or may
not possess solution, or possess nonunique solutions.
Oppositely, if this matrix is singular and the respective zero-mode
does not generate a new constraint, the system has a symmetry.

After this brief introduction, the SEM will be systematized. The main idea of this embedding formalism is to introduce extra fields into the model in order to obstruct the solutions of the Hamiltonian equations of motion.
It begins with the introduction of two arbitrary functions dependent on the original phase space and of WZ's variables, namely, $\Psi(a_i,p_i)$ and $G(a_i,p_i,\eta)$, into the first-order Lagrangian one-form as follows
\be
\label{2060a}
{\tilde{\cal L}}^{(0)}dt = A^{(0)}_\theta d\xi^{(0)\theta} + \Psi d\eta - {\tilde V}^{(0)}(\xi)dt,
\ee
with
\be
\label{2060b}
{\tilde V}^{(0)} = V^{(0)} + G(a_i,p_i,\eta),
\ee
where the arbitrary function $G(a_i,p_i,\eta)$ is expressed as an expansion in terms of the WZ field, given by
\begin{equation}
\label{2060}
G(a_i,p_i,\eta)=\sum_{n=1}^\infty{\cal G}^{(n)}(a_i,p_i,\eta),\,\,\,\,\,\,\,{\cal G}^{(n)}(a_i,p_i,\eta)\sim\eta^n\,,
\end{equation}
and satisfies the following boundary condition
\begin{eqnarray}
\label{2070}
G(a_i,p_i,\eta=0) = 0.
\end{eqnarray}
The symplectic variables were extended to also contain the WZ variable $\tilde\xi^{(0)\tilde\theta} = (\xi^{(0)\theta},\eta)$ (with ${\tilde\theta}=1,2,\dots,2N+1$) and the first-iterative symplectic potential becomes
\begin{equation}
\label{2075}
{\tilde V}^{(0)}(a_i,p_i,\eta) = V^{(0)}(a_i,p_i) + \sum_{n=1}^\infty{\cal G}^{(n)}(a_i,p_i,\eta).
\end{equation}
In this context, the new canonical momenta are
\be
{\tilde A}_{\tilde\theta}^{(0)} = \left\{\begin{array}{ll}
                                  A_{\theta}^{(0)}, & \mbox{with $\tilde\theta$ =1,2,\dots,2N}\\
                                  \Psi, & \mbox{with ${\tilde\theta}$= 2N + 1}
                                    \end{array}
                                  \right.
\ee
and the new symplectic tensor, given by
\begin{equation}
{\tilde f}_{\tilde\theta\tilde\beta}^{(0)} = \frac {\partial {\tilde A}_{\tilde\beta}^{(0)}}{\partial \tilde\xi^{(0)\tilde\theta}} - \frac {\partial {\tilde A}_{\tilde\theta}^{(0)}}{\partial \tilde\xi^{(0)\tilde\beta}},
\end{equation}
that is
\be
\label{2076b}
{\tilde f}_{\tilde\theta\tilde\beta}^{(0)} = \pmatrix{ { f}_{\theta\beta}^{(0)} & { f}_{\theta\eta}^{(0)}
\cr { f}_{\eta\beta}^{(0)} & 0}.
\ee

The implementation of the symplectic embedding scheme follows with two steps: the first one is addressed to compute $\Psi(a_i,p_i)$ while the second one is dedicated to the calculation of $G(a_i,p_i,\eta)$. In order to begin with the first step, we impose that this new symplectic tensor (${\tilde f}^{(0)}$) has a zero-mode $\tilde\nu$, consequently, we get the following condition
\begin{equation}
\label{2076}
\tilde\nu^{(0)\tilde\theta}{\tilde f}^{(0)}_{\tilde\theta\tilde\beta} = 0.
\end{equation}
Note that, at this point, $f$ becomes degenerate and, in consequence, we introduce an obstruction to solve, in an unique way, the Hamilton equation of motion given in Eq.(\ref{2050a1}). Assuming that the zero-mode $\tilde\nu^{(0)\tilde\theta}$ is
\begin{equation}
\label{2076a}
\tilde\nu^{(0)}=\pmatrix{\mu^\theta & 1},
\end{equation}
and using the relation given in (\ref{2076}) together with (\ref{2076b}), we get a set of equations, namely,
\be
\label{2076c}
\mu^\theta{ f}_{\theta\beta}^{(0)} + { f}_{\eta\beta}^{(0)} = 0,
\ee
where
\be
{ f}_{\eta\beta}^{(0)} =  \frac {\partial A_\beta^{(0)}}{\partial \eta} - \frac {\partial \Psi}{\partial \xi^{(0)\beta}}.
\ee
Observe that the matrix elements $\mu^\theta$ are chosen in order to disclose a desired gauge symmetry. Note that in this formalism the zero-mode $\tilde\nu^{(0)\tilde\theta}$ is the gauge symmetry generator. At this point, it is worth to mention that this characteristic is important because it opens up the possibility to disclose the desired hidden gauge symmetry from the noninvariant model. It awards to the symplectic embedding formalism some power to deal with noninvariant systems. From relation (\ref{2076}) some differential equations involving $\Psi(a_i,p_i)$ are obtained, Eq. (\ref{2076c}), and after a straightforward computation, $\Psi(a_i,p_i)$ can be determined.

In order to compute $G(a_i,p_i,\eta)$ in the second step, we impose that no more constraints arise from the contraction of the zero-mode $(\tilde\nu^{(0)\tilde\theta})$ with the gradient of the potential ${\tilde V}^{(0)}(a_i,p_i,\eta)$. This condition generates a general differential equation, which reads as
\begin{widetext}
\begin{eqnarray}
\label{2080}
\tilde\nu^{(0)\tilde\theta}\frac{\partial {\tilde V}^{(0)}(a_i,p_i,\eta)}{\partial{\tilde\xi}^{(0)\tilde\theta}}\,&=&\, 0,\\
\mu^\theta \frac{\partial {V}^{(0)}(a_i,p_i)}{\partial{\xi}^{(0)\theta}} + \mu^\theta \frac{\partial {\cal G}^{(1)}(a_i,p_i,\eta)}{\partial{\xi}^{(0)\theta}} 
\,+\, \mu^\theta\frac{\partial {\cal G}^{(2)}(a_i,p_i,\eta)}{\partial{\xi}^{(0)\theta}} + &\dots& 
\,+\,\frac{\partial {\cal G}^{(1)}(a_i,p_i,\eta)}{\partial\eta} + \frac{\partial {\cal G}^{(2)}(a_i,p_i,\eta)}{\partial\eta} + \dots = 0\;\;, \nonumber
\end{eqnarray}
\end{widetext}
that allows us to compute all correction terms ${\cal G}^{(n)}(a_i,p_i,\eta)$ in order of $\eta$. Note that this polynomial expansion in terms of $\eta$ is equal to zero, consequently, whole coefficients for each order in $\eta$ must be null identically. In view of this, each correction term in order of $\eta$ is determined. For a linear correction term, we have
\begin{equation}
\label{2090}
\mu^\theta\frac{\partial V^{(0)}(a_i,p_i)}{\partial\xi^{(0)\theta}} + \frac{\partial{\cal
 G}^{(1)}(a_i,p_i,\eta)}{\partial\eta} = 0.
\end{equation}
For a quadratic correction term, we get
\begin{equation}
\label{2095}
{\mu}^{\theta}\frac{\partial{\cal G}^{(1)}(a_i,p_i,\eta)}{\partial{\xi}^{(0)\theta}} + \frac{\partial{\cal G}^{(2)}(a_i,p_i,\eta)}{\partial\eta} = 0.
\end{equation}
From these equations, a recursive equation for $n\geq 2$ is proposed as
\begin{equation}
\label{2100}
{\mu}^{\theta}\frac{\partial {\cal G}^{(n - 1)}(a_i,p_i,\eta)}{\partial{\xi}^{(0)\theta}} + \frac{\partial{\cal
 G}^{(n)}(a_i,p_i,\eta)}{\partial\eta} = 0,
\end{equation}
that allows us to compute the remaining correction terms in order of $\eta$. This iterative process is successively repeated until (\ref{2080}) becomes identically null, consequently, the extra term $G(a_i,p_i,\eta)$ is obtained explicitly. Then, the gauge invariant Hamiltonian, identified as being the symplectic potential, is obtained as
\begin{equation}
\label{2110}
{\tilde{\cal  H}}(a_i,p_i,\eta) = V^{(0)}(a_i,p_i) + G(a_i,p_i,\eta),
\end{equation}
and the zero-mode ${\tilde\nu}^{(0)\tilde\theta}$ is identified as being the generator of an infinitesimal gauge transformation, given by
\begin{equation}
\label{2120}
\delta{\tilde\xi}^{\tilde\theta} = \varepsilon{\tilde\nu}^{(0)\tilde\theta},
\end{equation}
where $\varepsilon$ is an infinitesimal parameter.

\section{The Symplectic analysis}

The study of both Lorentz and gauge invariance in variations of
Maxwell's model is of strong theoretical
\cite{Colladay,Colladay2,Coleman,Coleman2,Jackiw,Jackiw2,Andrianov,Andrianov2,Adam,Perez,Perez2,Baeta,Baeta2} and
experimental \cite{Carroll} interest and great relevance in
practical applications as the quantum Hall effect \cite{Girvin,Girvin2,Girvin3}
and high-$T_c$ superconductivity \cite{Polyakov,Polyakov2}.

In this section, the MCS theory in four dimensions will be
analyzed from the symplectic point of view. Let us consider the
massive Maxwell-Chern-Simons Lagrangian in four dimensions
\cite{Carroll,helayel}
\be \label{01} {\cal L}=-\frac{\beta}{4}F_{\mu\nu}F^{\mu\nu}
+\frac{m^2}{2}A_\mu A^\mu -\frac{1}{4}p_\alpha A_\beta
\epsilon^{\alpha\beta\mu\nu}F_{\mu \nu}, \ee
where $p$ is an external four-vector. Now, following the
symplectic method the zeroth-iterative first-order Lagrangian
one-form is written as
\begin{widetext}
\ba \label{02} 
{\cal L}\,=\,\pi^i \dot A_i &+&A_0\left( \partial^i
\pi_i +m^2 A_0 +\frac{1}{4}p^i\epsilon_{ijk} F^{ik}\right)
\,+\,\frac{1}{2\beta}p_i A_j \pi_k \epsilon^{ijk} 
+\frac{1}{2\beta}\pi_i \pi^i
\,-\,\frac{\beta}{4}F_{ij}F^{ij} -\frac{1}{2}m^2 A_0 A^0 
\,+\,\frac{1}{2}m^2A_iA^i \nonumber \\
&+&\frac{1}{8\beta}p^iA_j \(p_jA^i -p_i A^j \) 
\,-\, \frac{1}{4}p^0 A^i \epsilon_{ijk}F^{jk} \;\;,
\ea
\end{widetext}
with the canonical momentum $\pi_i$  given by
\ba \label{03} \pi_i &=& -\beta F_{0i} -\frac{1}{2}p^jA^k
\epsilon_{ijk}\nonumber\\
&=& -\beta\left(\dot A_i -\partial_i A_0\right) -\frac{1}{2}p^jA^k
\epsilon_{ijk}. \ea

The symplectic fields are $\xi^{(0)\alpha}\,=\,\(A^i ,\pi^i ,A^0 \)$
and the zeroth-iterative symplectic matrix is
\be \label{04} f^{(0)}=
\left(%
\begin{array}{ccc}
  0 & -\delta^i_j & 0 \\
  \delta^j_i & 0 & 0 \\
  0 & 0 & 0 \\
\end{array}%
\right)\delta(x-y) \ee
which is a singular matrix. It has a zero-mode that generates the
following constraint
\be \label{05} \Omega(x) = \partial_i \pi^i(x) + m^2 A_0 (x)
+\frac{1}{4}p^iF^{jk}\epsilon_{ijk}, \ee
identified as being the Gauss law. Bringing back this constraint
into the canonical part of the first-order Lagrangian density
${\cal L}^{(0)}$ using a Lagrangian multiplier $\(\zeta\)$, the
first-iterated Lagrangian density, written in terms of the following
symplectic fields $\xi^{(1)\alpha} =\(A^i,\pi^i,A^0,\zeta\)$ is
obtained as
\begin{widetext}
\ba \label{06} 
{\cal L}^{(1)} &=&\pi^i \dot A_i +\Omega\dot \zeta
\,+\,\frac{1}{2\beta}p_i A_j \pi_k \epsilon^{ijk}
\,+\,\frac{1}{2\beta}\pi_i \pi^i 
\,-\,\frac{\beta}{4}F_{ij}F^{ij}
-\frac{1}{2}m^2 A_0 A^0 \,+\,\frac{1}{2}m^2A_iA^i \nonumber\\
&+&\frac{1}{8\beta}p^iA_j \(p_jA^i -p_i A^j \) 
\,-\, \frac{1}{4}p^0 A^i \epsilon_{ijk}F^{jk}. 
\ea
\end{widetext}
The first-iterated symplectic matrix is obtained as being
\ba 
& &\label{07} f^{(1)}= \nonumber\\
& & \!\!\!\!\!\!\!\left(
\begin{array}{cccc}
  0 & -\delta_j^i \delta(x-y) & 0 & f_{A^i \zeta} \\
  \delta_i^j\delta(x-y) & 0 & 0 & \partial_i^y\delta(x-y) \\
  0 & 0 & 0 & m^2\delta(x-y) \\
  f_{\zeta A^j} & -\partial_j^y\delta(x-y) & -m^2\delta(x-y) & 0 \\
\end{array}\right) \nonumber\\
\mbox{}
\ea
where
\be \label{08} f_{A^i \zeta}=-\frac{1}{2}p^n\partial_y^m
\delta(x-y) \epsilon_{nim}. \ee
This matrix is nonsingular and, as settle by the symplectic
formalism, the Dirac brackets among the phase space fields are
acquired from the inverse of the symplectic matrix, namely,
\ba \label{09} \{A^i(x),A^j(y)\}^*&=&0,\nonumber\\
\{A^i(x),\pi^j(y)\}^* &=& \delta^{ij}\delta(x-y),\nonumber\\
\{A^0(x),A^j(y)\}^* &=& \frac{1}{m^2}\partial_x^j \delta(x-y),\\
\{A^0(x),\pi^j(y)\}^* &=&
\frac{1}{2m^2}\epsilon^{lij}p_l\partial_i^x\delta(x-y).\nonumber
\ea

As we said above, the basic symplectic analysis was the first step of the SEM.  The next step is to introduce the WZ fields in order to proceed with the dualization.  This will be done in the next section.

\section{The Dual Equivalent Model}

Now the phase space will be extended with the introduction of the WZ
fields. In order to start, we change the Lagrangian,
Eq. (\ref{02}), introducing two arbitrary functions
$\psi\equiv\psi\(A^i,\pi^i,A^0, \eta\)$ and $G\equiv
G\(A^i,\pi^i,A^0, \eta\)$ with the WZ field, namely,
\be \label{11} \tilde{\cal L}^{(0)} =\pi_i \dot A^i + \psi\dot\eta
-\tilde{V}^{(0)}, \ee
where the symplectic potential is
\begin{widetext}
\ba \label{12} 
\tilde {V}^{(0)} &=&-A_0\left( \partial^i \pi_i
+m^2 A_0 \,+\,\frac{1}{4}p^i\epsilon_{ijk} F^{ik}\right) 
\,-\,\frac{1}{2\beta}p_i A_j \pi_k \epsilon^{ijk}
-\frac{1}{2\beta}\pi_i \pi^i +\frac{\beta}{4}F_{ij}F^{ij}
\,+\,\frac{1}{2}m^2 A_0 A^0 -\frac{1}{2}m^2A_iA^i \nonumber \\
&-&\,\frac{1}{8\beta}p^iA_j \(p_jA^i -p_i A^j \) 
\,+\, \frac{1}{4}p^0 A^i \epsilon_{ijk}F^{jk} \,+\, G, 
\ea
\end{widetext}
and $G$ is a function expressed as
\be \label{13} G(A^i,\pi^i, A^0, \eta)=\sum_{n=1}^\infty {\cal
G}^{n} \,\,\, \mbox{with} \,\,\, {\cal G}^{n} \propto\eta^n .\ee
The arbitrary function satisfies the following boundary condition,
\be \label{14} G\(A^i,\pi^i,A^0,\eta=0 \) =0. \ee

The extended symplectic field are $\tilde \xi^{(0)}
=\(A^i,\pi^i,A^0,\eta\)$ and the corresponding matrix is
\be\label{15} \tilde{f}^{(0)}=\left(\begin{array}{cccc}
  0 & -\delta_j^i\delta(x-y) & 0 & \frac{\delta\psi(y)}{\delta A^i(x)} \\
  \delta_i^j\delta(x-y) & 0 & 0 & \frac{\delta\psi(y)}{\delta \pi^i(x)} \\
  0 & 0 & 0 & \frac{\delta\psi(y)}{\delta A^0(x)} \\
  -\frac{\delta\psi(x)}{\delta A^j(y)} & -\frac{\delta\psi(x)}{\delta \pi^j(y)} & -\frac{\delta\psi(x)}{\delta A^0(y)} & 0 \\
\end{array}
\right). \ee
This singular matrix has a zero-mode, which can be settle conveniently as
\be \label{16} \tilde \nu = \pmatrix{\partial^i & 0 & \partial^0 &
1 }. \ee
Contracting this zero-mode with the symplectic matrix above, a set
of differential equation can be obtained as
\ba \label{17} 
& &\int dx\;\; \( \frac{\delta\psi(y)}{\delta
A^i(x)}\)\,=\,0, \nonumber\\
& &\int dx\;\; \( \delta_i^j\partial^x_j\delta(x-y) +
\frac{\delta\psi(y)}{\delta\pi^i(x)}\)\,=\,0,\nonumber\\
& &\int dx\;\; \( \frac{\delta\psi(y)}{\delta A^0(x)}\)\,=\,0, \\
& &\int dx\;\; \(-\partial_x^j\frac{\delta\psi(x)}{\delta A^j(y)}
-\partial_x^0\frac{\delta\psi(x)}{\delta A^0(y)}\)\,=\,0. \nonumber
\ea
After a straightforward computation, we get
\be \label{18} \psi(x) = -\partial^i \pi_i(x). \ee
Then, the Lagrangian becomes
\be \label{19} \tilde{\cal L}^{(0)} =\pi_i \dot A^i
-\partial^i\psi_i\dot\eta -\tilde{V}^{(0)}\,. \ee

After this point, we begin with the final step of the symplectic
embedding formalism. To this end, we impose that the contraction
of the zero-mode, Eq. (\ref{16}), with the gradient of the
symplectic potential generates an identically null result, namely,
\be\label{20} \int dy\;\; \tilde \nu^{(0)}(x)\frac{\delta\tilde
V^{(0)}(y)}{\delta\tilde\xi^{(0)}(x)}=0\,\,. \ee
From this condition, the following general differential equation
is obtained,
\ba \label{21} 
& & \int dy \,\[\partial_x^i \(\frac{\delta\tilde{V}^{(0)}(y)}{\delta A^i(x)}\)+
\partial_x^0 \(\frac{\delta\tilde{V}^{(0)}(y)}{\delta A^0(x)}\) \right. \nonumber \\
& & \left.+\; 1.\(\sum_{n=1}^{\infty}\frac{\delta{\cal G}^{(n)}(y)}{\delta \eta(x)}\)
\]=0,
\ea
where the relation given in (\ref{13}) was used. This allows
the computation of the whole correction terms in order of $\eta$.
For linear correction term $\({\cal G}^{(1)}(x)\)$, we get
\begin{widetext}
\ba \label{22} 
{\cal G}^{(1)}\,&=&\,\[{1\over2}p^i\partial^jA^0(x)\epsilon_{ijl}
\,+\,{1\over2}p^iF^{0k}(x)\epsilon_{ilk}
-\beta\partial^iF_{il}(x)\,-\,m^2A_l(x)+
{1\over4}p^0F^{jk}(x)\epsilon_{jkl}-{1\over2}p^0\partial^j
A^i\epsilon_{ijl}\]\partial^l \eta \nonumber\\
&-&\[\partial^i\pi_i(x) +m^2A_0(x)
+{1\over4}p^iF^{jk}(x)\epsilon_{ijk}\]\partial^0\eta.  
\ea
\end{widetext}

For the quadratic correction term, we have
\ba \label{23} 
& &\int dy \[
\partial_x^i \(\frac{\delta{\cal G}^{(1)}(y)}{\delta A^i(x)}\)+
\partial_x^0 \(\frac{\delta{\cal G}^{(1)}(y)}{\delta A^0(x)}\) \right. \nonumber \\
& & \left.+\, 1.\(\frac{\delta{\cal G}^{(2)}(y)}{\delta \eta(x)}\)
\]=0,
\ea
with the following solution,
\be\label{24}{\cal
G}^{(2)}=-{m^2\over2}\partial_i\eta\partial^i\eta-
{m^2\over2}\partial_0\eta\partial^0\eta. \ee

Note that the second-order correction term has dependence only on the
WZ field, thus all the correction terms ${\cal G}^{(n)}$ for
$n\geq 3$ are null. Then, the gauge invariant first-order
Lagrangian density is given by
\begin{widetext}
\ba \label{25} 
\tilde{\cal L} &=& {\cal L} \,+\,\[ m^2 A_k
+\beta\partial^0 F_{0k} +\beta\partial^i F_{ik} -p^0\partial^iA^j
\epsilon_{ijk} 
\,+\, p^iF^{j0}\epsilon_{ijk}\]\partial^k\eta\,+\,\[m^2 A_0 +\beta\partial^i F_{io}-
p^i\partial^jA^k\epsilon_{ijk}\]\partial^0\eta \nonumber\\
&+&{m^2\over2}\(\partial_i\eta\partial^i\eta
\,+\,\partial_0\eta\partial^0\eta\), 
\ea
\end{widetext}
where $\cal L$ is given in (\ref{01}). We may recognize the
Noether current in Eq.(\ref{25}) as
\ba\label{26}
J_k &=&m^2 A_k +\beta\partial^0 F_{0k}
+\beta\partial^i F_{ik} -p^0\partial^iA^j \epsilon_{ijk}
+p^iF^{j0}\epsilon_{ijk}, \nonumber\\
J_0 &=&m^2 A_0 +\beta\partial^i F_{io}-
p^i\partial^jA^k\epsilon_{ijk}. \ea
So, we can write $\tilde {\cal L}$ as
\be\label{27} \tilde{\cal L} = {\cal L} +J_{\mu}\partial^{\mu}\eta
+ {m^2\over2}\partial_{\mu}\eta\partial^{\mu}\eta . \ee
Solving for $\partial_\mu \eta$ we get that
\be\label{28} J_\mu +m^2\partial_{\mu}\eta=0. \ee
Plugging this back into (\ref{27}), we obtain the remarkable
gauge-invariant theory
\begin{widetext}
\ba\label{29} 
\tilde{\cal L} &=&{\beta\over4}F_{\mu\nu}F^{\mu\nu}+
{1\over2}\epsilon_{\alpha\beta\nu\mu}p^{\alpha}\(\partial^{\beta}A^{\nu}\)A^{\mu} 
\,-\,{1\over{2m^2}}\[\epsilon_{\alpha\beta\nu\mu}p^{\alpha}\(\partial^{\beta}A^{\nu}\)\]^2 
\,-\,{\beta^2\over{2m^2}} 
\[\partial_\mu F^{\mu\nu}\]^2 \nonumber\\
&+&{\beta\over{m^2}}\;\epsilon_{\alpha\beta\nu\mu}p^{\alpha}\(\partial^{\beta}A^{\alpha}\)\(\partial_{\rho}F^{\rho\mu}\).
\ea
\end{widetext}
which is the same result obtained in \cite{helayel}, using the NDM, i.e., the action (\ref{29}) is the dual equivalent to the action (41).  We see that as the zero-mode given in (\ref{16}) is arbitrary, any other zero-mode for the symplectic matrix (54) will bring us a new dual equivalent action.  We believe that this is one great advantage of the SEM when we confront this with the NDM.  Since our objective in this work is to prove that the SEM can produce dual equivalent actions as the NDM, we used the zero-mode which reproduce the action obtained in [26], namely, the Eq. (\ref{29}).

To complete the comparison between both methods, as well known from the symplectic formalism literature, the zero-mode is the generator of the infinitesimal gauge transformations, since Eq. (\ref{29}) is also the gauge invariant versions of (41).  We believe that this constitute another good point in favor for the SEM.  So, using the zero mode,
Eq. (\ref{16}), as the generator of the infinitesimal gauge
transformations given by  $\delta{\cal O}=\epsilon \tilde \nu^{(0)}$, we have
\ba\label{30} \delta A_i &=& -\partial_i \epsilon,\nonumber\\
\delta \pi_i &=&0,\nonumber\\
\delta A_0 &=& -\partial_0 \epsilon, \\
\delta \eta &=& \epsilon, \nonumber \ea
where $\epsilon$ is an infinitesimal time-dependent parameter.

\section{Conclusions}

The investigation of how to obtain dual equivalent actions to systems with many degrees of freedom is quite desirable since these systems live in a world permeated with non-perturbative features that need special and difficult treatment.

The technique of symplectic embedding follows the idea of Faddeev and Shatahvilli and is based on a contemporary framework that handles constrained models, namely, the symplectic formalism.  The effectiveness of the method was demonstrated through several papers in the literature and the positive points in favor of it are many, as we described in the introduction.

We believe that the objective reached by this paper was to prove another positive point of SEM, which was to surpass the recently developed NDM method since SEM show a deeper insight from the moment that a whole family of dual equivalent actions can be disclosed.  This can be easily realized in the arbitrariness of the zero-mode, which also results in another advantageous point since the zero-mode is the generator of the infinitesimal gauge transformation (Eq. (\ref{30})).  We think that these properties compensate the additional calculation performed in SEM in comparison with NDM.

To exemplify our conclusions we promote the dualization of the gauge-invariant Maxwell theory modified by the
introduction of an explicit massive (Proca) term and a
topological but not Lorentz-invariant term \cite{Carroll,helayel}.   Afterwards, this
noninvariant theory was reformulated as a gauge-invariant/dual equivalent theory
via SEM where the gauge-invariance broken was restored. This result reproduces the version of the
theory obtained in \cite{helayel} via NDM.

\section{ Acknowledgments}

EMCA would like to thank the hospitality and kindness of the Dept. of Physics of the Federal University of Juiz de Fora where part of this work was done.
The authors would like  to thank CNPq, FAPEMIG and FAPERJ (Brazilian financial agencies) for financial support.


\begin{thebibliography} {99}


\bibitem{ainrw} M. A. Anacleto, A. Ilha, J. R. S. Nascimento, R. F. Ribeiro and C. Wotzasek, Phys. Lett. B 504, 268 (2001).

\bibitem{iw}  A. Ilha and C. Wotzasek, Nucl. Phys. B 604, 426 (2001).

\bibitem{iw2}  A. Ilha and C. Wotzasek, Phys. Lett. B 519, 169  (2001). 

\bibitem{iw3}  A. Ilha and C. Wotzasek, Phys. Lett.B 510, 329  (2001). 

\bibitem{tpn} P. K. Townsend, K. Pilch and P. van Nieuwenhuizen, Phys. Lett. B 136, 38 (1984).

\bibitem{dj} S. Deser and R. Jackiw, Phys. Lett. B 139, 2366 (1984).

\bibitem{suecos}  S. E. Hjelmeland, U. Lindstr\"om, UIO-PHYS-97-03, May 1997, e-Print Archive: hep-th/9705122.

\bibitem{dirac}  P. A. M. Dirac, Proc. R. Soc. London A257 (1960) 32; {\it Lectures on Quantum Mechanics} (Yeshiva University Press, New York, 1964). 

\bibitem{dirac2}  A. Hanson, T. Regge and C. Teitelboim, {\it Constrained Hamiltonian Systems} (Academia Nazionale dei Lincei, Rome, 1976). 

\bibitem{dirac3}  K. Sundermeyer, {\it Constrained Dynamics}, Lectures Notes in Physics, Vol. 169 (Springer, New York, 1982). 

\bibitem{dirac4}  P. G. Bergmann, Phys. Rev. 75, 680 (1949). 

\bibitem{dirac5}  P. G. Bergmann, Phys. Rev. 89, 4 (1953).

\bibitem{FS}  L.Faddeev and S.L.Shatashivilli, Phys. Lett. B 167, 225 (1986).

\bibitem{batalin}   I. A. Batalin and E. S. Fradkin, Nucl. Phys. B 279, 514 (1987).  

\bibitem{batalin2}   I. A. Batalin and I. V. Tyutin, Int. J. Mod. Phys. A 6, 3255 (1991).

\bibitem{wotzasek}   C. Wotzasek, Int. J. Mod. Phys. A 5, 1123 (1990). 

\bibitem{wotzasek2}   C. Wotzasek, Phys. Rev. Lett. 66, 129 (1991).

\bibitem{mo}  M. Moshe and Y. Oz, Phys. Lett. B 224, 224 (1989). 

\bibitem{mo2}  T. Fujiwara, Y. Igarashi and J. Kubo, Nucl. Phys. B 341, 695 (1990).

\bibitem{barcelos}  J. Barcelos-Neto, Phys. Rev. D 55, 2265 (1997).

\bibitem{soldati} A. A. Andrianov and R. Soldati, Phys. Lett. B 435, 449 (1998).

\bibitem{klink} C. Adam and F.R. Klinkhamer, Nucl. Phys. B 607, 247 (2001).

\bibitem{nodland} B. Nodland and J. P. Ralston, Phys. Rev. Lett. 78, 3043 (1997).

\bibitem{ggw}   M. S. Guimaraes, L. Grigorio, C. Wotzasek, hep-th/0609215.

\bibitem{cohen} J. F. L. Wardle, R. A. Perley and M. H. Cohen, Phys. Rev. Lett. 79, 1801 (1997).

\bibitem{aon}   J. Ananias Neto,  C. Neves and W. Oliveira, Phys. Rev. D 63, 085018 (2001).

\bibitem{vyth}   A. S. Vytheeswaran, Int. J. Mod. Phys. A 13, 765 (1998).

\bibitem{mitra}   P. Mitra, Ann. Phys. (N.Y.) 203, 137 (1998).

\bibitem{FJ}   L. Faddeev and R. Jackiw, Phys. Rev. Lett. 60, 1692 (1988).

\bibitem{FJ2}    N. M. J. Woodhouse, {\it Geometric Quantization}, Clarendon Press, Oxford 1980.

\bibitem{FJ3}   J. Barcelos Neto and C. Wotzasek, Mod. Phys. Lett. A 7, 1737 (1992). 

\bibitem{FJ4}   J. Barcelos Neto and C. Wotzasek, Int. J. Mod. Phys. A 167, 4981 (1992).

\bibitem{helayel}  M. Botta Cantcheff, C.F.L. Godinho, A.P. Ba\^eta Scarpelli and J.A. Hellay\"el-Neto, Phys. Rev. D 68, 065025 (2003).

\bibitem{fluidos}   A. C. R. Mendes, C. Neves and W. Oliveira, J. Phys. A 37, 1927 (2004).

\bibitem{henrique}  C. P. Natividade and H. Boschi-Filho, Phys. Rev. D 62, 025016 (2000).

\bibitem{bazeia}  D. Bazeia and R. Jackiw, Ann. Phys. 270, 246 (1998).

\bibitem{br}   R. Banerjee and H. J. Rothe, Nucl. Phys. B 447, 183 (1995).

\bibitem{gotay}  M. J. Gotay, J. M. Nester and G. Hinds, J. Math. Phys. 19(11), 2388 (1978).

\bibitem{Colladay}  D. Colladay and V. A. Kosteleck\'y, Phys. Rev. D 55, 6760 (1998).

\bibitem{Colladay2}  D. Colladay and V. A. Kosteleck\'y,  Phys. Rev. D 55, 6760 (1997).

\bibitem{Coleman}  S. Coleman and S. L. Glashow, Phys. Lett B 405, 249 (1997). 

\bibitem{Coleman2}  S. Coleman and S. L. Glashow,  Phys. Rev. D 59, 116008 (1990).

\bibitem{Jackiw}  R. Jackiw and V. A. Kosteleck\'y, Phys. Rev. Lett. 82, 3572 (1999).

\bibitem{Jackiw2}  R. Jackiw and S. Templeton, Phys. Rev. D 23, 2291 (1981).

\bibitem{Andrianov}   A. A. Andrianov and R. Soldati, Phys. Rev. D 51, 5961 (1995).

\bibitem{Andrianov2}   A. A. Andrianov and R. Soldati, Phys. Lett. B 435, 449 (1998).

\bibitem{Adam}   C. Adam and F. R. Klinkhamer, Nucl. Phys. B 607, 247 (2001).

\bibitem{Perez}   M. P\'erez-Victoria, Phys. Rev. Lett. 83, 2518 (1999).

\bibitem{Perez2}   M. P\'erez-Victoria,  J. High Energy Phys. 04, 032 (2001).
 
\bibitem{Baeta}  P. Ba\^eta Scarpelli, M. Sampaio and M.C. Nemes, Phys. Rev. D 63, 046004 (2001).

\bibitem{Baeta2}   A.P. Ba\^eta Scarpelli, M. Sampaio, B. Hiller and M. C. Nemes, Phys. Rev. D 64, 046013 (2001).

\bibitem{Carroll}  Sean M. Carroll, George B. Field and R. Jackiw, Phys. Rev. D 41, 1231 (1990).

\bibitem{Girvin}  S. Girvin and R. Prange, {\cal The Quantum Hall Effect}(Springer, Berlin, 1987).

\bibitem{Girvin2}   Z. Zhang, T. Hansson and S. Kivelson, Phys. Rev. Lett. 62, 82 (1989).

\bibitem{Girvin3}   Z. Zhang, T. Hansson and S. Kivelson, Phys. Rev. Lett.  62, 980(E) (1989).

\bibitem{Polyakov}   A. Polyakov, Mod. Phys. Lett. A 3, 325 (1988).

\bibitem{Polyakov2}   Y. H. Chen, F. Wilezek, E. Witten and B. Halperin, Int. J. Mod. Phys. B 3, 1001 (1989).


\end{thebibliography}
\end{document}